\documentclass{epl}

\title{On the mechanism of irradiation-enhanced exchange bias}

\author{S. Poppe \and J. Fassbender \and B. Hillebrands}
\shortauthor{S. Poppe \etal}

\institute{
  Fachbereich Physik and Forschungsschwerpunkt MINAS, Technische Universit{\"a}t Kaiserslautern -
           Erwin-Schr\"{o}dinger-Stra{\ss}e 56, 67663 Kaiserslautern, Germany
}

\pacs{75.70.Cn}{Magnetic properties of interfaces}
\pacs{61.80.Jh}{Ion irradiation effects} \pacs{75.30.Gw}{Magnetic
anisotropy }

\begin{document}

\maketitle

\begin{abstract}

By means of layer resolved ion irradiation the mechanisms
involved in the irradiation driven modifications of the exchange
bias effect in NiFe/FeMn bilayers have been investigated. It is
shown that not only the locations of the defects but also the
magnetic coupling between both layers during the irradiation
process is of crucial importance. This requires an extension of
current models accounting for defects in exchange bias systems.

\end{abstract}

\section{Introduction}

The magnetic exchange interaction between an antiferromagnetic (AF) and a
ferromagnetic (F) layer can lead to the so called exchange bias effect
\cite{Me56}. The most prominent feature is a shift of the hysteresis loop
along the field axis. It is widely exploited in applications like angular
sensors and magnetic memory cells to provide a fixed reference direction in
one of the two magnetic layers of a magnetoresistive device.

Several models have been proposed that account for various related phenomena
\cite{Mal87,Koon99,Sc99,Ki99,Now02}, taking into account domains in the AF
layer either parallel or perpendicular to the layer plane. Though a complete
understanding of the effect on the microscopic scale is still under
discussion, it is generally agreed upon that exchange bias is very sensitive
to the structural properties of the AF layer and its interface to the F layer.
A good demonstration of this sensitivity has been performed either by diluting
the AF layer with non-magnetic atoms \cite{Milt,Shi} or by ion irradiation of
the whole layer system with light ions of several keV energy \cite{PRB}. In
both cases an enhancement of the exchange bias field over the initial value
was observed. Additionally, it has been demonstrated that irradiation can be
used to adjust the pinning direction of the exchange bias effect \cite{PRB}.

In order to understand the ion dose dependence of the exchange bias field a
phenomenological model has been proposed \cite{PRB}. It takes into account
structural effects in the AF layer, similar to the suggestions in
Ref.~\cite{Milt}, as well as at its interface with the F layer.

In the present work we focus on the mechanisms responsible for the
change of the exchange bias field magnitude during ion
irradiation. By performing the irradiation process between
deposition steps without breaking the UHV environment, the effects
of bombardment are limited to specific regions of the layer stack.
Thus, a first experimental proof of the role of the defect
position within the bilayer as suggested in the proposed model
\cite{PRB} is provided. In addition to this model, irradiation
experiments at elevated temperatures demonstrate that the
enhancement of the exchange bias field is driven by the magnetic
coupling between the layers. An alternative approach to the
modification mechanisms is suggested.

Before going into the details of the experiments, we will provide
a short summary of the model described in Ref.~\cite{PRB}. It is
assumed that the ion induced modifications have different effects
on the exchange bias depending on their vertical placement within
the layer stack. Defects created in the volume of the AF layer are
supposed to act as pinning sites for AF domain walls. These
pinning sites reduce the energy necessary to create new domain
walls, thus increasing the density of walls upon irradiation and
decreasing the average size of the domains. According to
random-field models this leads to a larger shift field, see, e.g.,
Ref.~\cite{Mal87}. In first approximation, this is described by
the relation $H_{eb}(n) \propto (1+aptn)$, with $n$ the ion dose,
$t$ the AF layer thickness, $p$ the efficiency of defect creation
and $a$ describing the efficiency of a volume defect to change the
bias field. $p$ is calculated using the SRIM code \cite{TRIM}.

Second, the ion bombardment leads to a mixing of the AF/F
interface which will suppress the exchange bias due to broken
exchange interaction across the interface. This is described by an
exponential decay with dose: $H_{eb}(n)\propto \exp(-b_I n)$, with
$b_I$ describing the efficiency of an interface defect to change
the bias field. The total dose dependence therefore is given by
\begin{equation}\label{dosedepeq}
H_{eb}(n) / H_{eb}(0) = (1+aptn) \cdot \exp(-b_I n).
\end{equation}
For a more detailed description see Ref.~\cite{PRB}.

\section{Experiments}
The studied samples are polycrystalline and have been prepared by
thermal evaporation in a UHV system with a base pressure of $5
\cdot 10^{-10}$\,mbar. A 150\,{\AA} thick Cu buffer layer has been
deposited on top of an oxidized Si substrate. For the F (AF) layer
50\,{\AA} (100\,{\AA}) of Ni$_{81}$Fe$_{19}$ (Fe$_{50}$Mn$_{50}$)
has been used. The stacking sequence of these two layers has no
significant effect on the observed exchange bias or the dose
dependence after irradiation. Finally the sample is covered with
20\,{\AA} of Cr to protect it from oxidation. For the ion
bombardment a commercial sputter gun has been used which is
attached to the same vacuum system and is operated with 5\,keV He
ions. The base pressure during irradiation is $5\cdot
10^{-8}$\,mbar. The sample is exposed to different ion doses at
several areas by scanning the ion beam and changing exposure time.
The ion current is controlled by a Faraday cup. Transmission
electron microscopy (TEM) studies have been performed to check for
changes in microstructure and texture after irradiation of the
whole layer stack. At a dose of $2 \cdot 10^{15}$ ions/cm$^2$
giving the maximum enhancement no changes have been detected.

\begin{figure}
\onefigure{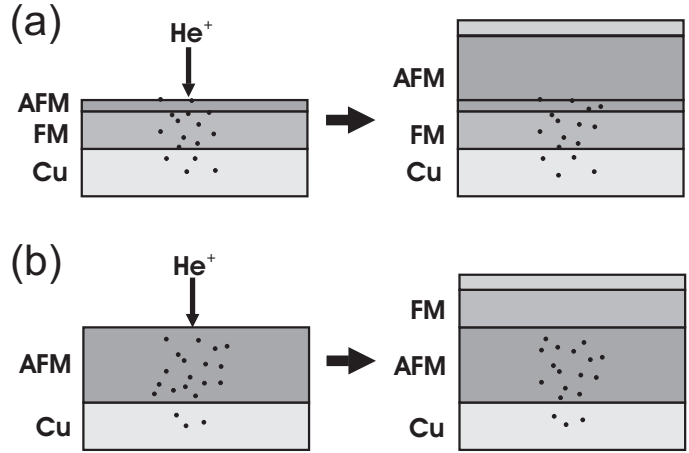}
\caption{Scheme of the experiments for investigation of defects in (a) the
interface region only, (b) the volume AF layer only.} \label{Schema}
\end{figure}

\begin{figure}
\onefigure{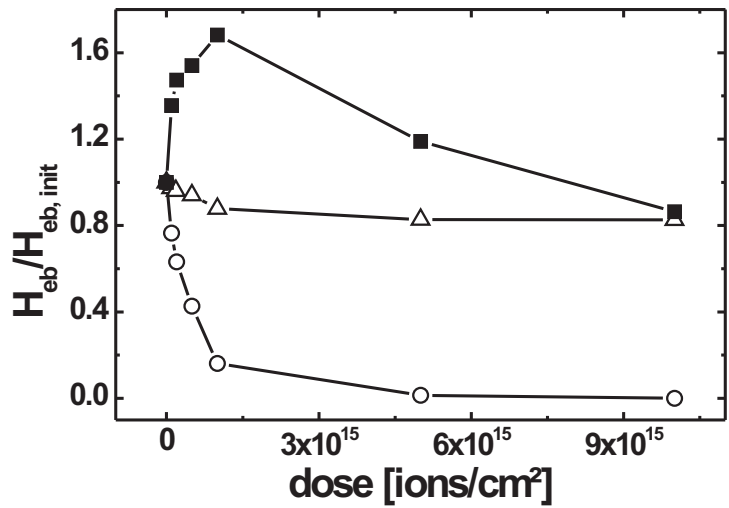}
\caption{Dose dependence of H$_{eb}$ normalized to the as-prepared value. The
closed squares correspond to a sample that has been irradiated completely. The
open circles belong to a sample where only the interface region has been
modified, and triangles denote selective irradiation of the AF layer.}
\label{Insitu}
\end{figure}

In the first experiment, the influence of ion bombardment on the
AF/F interface was studied. For this purpose, a sample was
prepared with the F layer grown before the AF layer. Growth of the
AF layer was stopped at a thickness of 15\,\AA. At this thickness
the interface is well defined but the sample exhibits neither a
bias shift nor an enhancement of coercivity yet. Several areas
with ion doses ranging from $9 \cdot 10^{13}$ to $2 \cdot
10^{16}$\,ions/cm$^{2}$ were irradiated. During the bombardment a
field of 30\,Oe was applied to saturate the F layer. Next, the
deposition of the AF layer was completed to obtain a total
thickness of 100\,\AA. A scheme of the experiment is shown in
Fig.~\ref{Schema}. After growth the sample was heated above the
blocking temperature to 200$^\circ$C and cooled in an applied
field of 250\,Oe to initialize the exchange bias. Magnetic
characterization was performed by longitudinal magneto-optical
Kerr effect (MOKE) magnetometry. It should be stated that the
absolute value of $H_{eb}$ at zero dose is reduced to 80\,Oe in
this sample compared to typical values of 180\,Oe in cases where
growth of FeMn has not been interrupted. This is probably due to a
residual contamination of the surface during irradiation reducing
the overall coupling. However, this value is still one order of
magnitude larger than anisotropies which have been reported in
single Ni$_{81}$Fe$_{19}$ films \cite{Woods}. Therefore, this does
not affect the conclusions drawn in the following. In
Fig.~\ref{Insitu} the dose dependence of the normalized exchange
bias field compared to a sample which has been irradiated after
completed deposition is shown. An enhancement of the exchange bias
field is only observed for the sample where all layers have been
irradiated. In the case where the irradiation can only cause
intermixing an enhancement is completely absent. This is the
behaviour as expected from the model.

In a complementary experiment the effect of bombardment on the volume AF layer
was studied. Therefore, the AF layer was grown first and irradiated with the
same dose pattern as described above. Then the NiFe layer and the Cr
protective layer were deposited. The sample was annealed under the same
conditions as described above.  It exhibits a shift field of 180\,Oe on the
non-irradiated areas. The exchange bias field has only a weak dependence on
the ion dose in this case (Fig.~\ref{Insitu}, triangles). Especially, it does
not decay to zero like it is observed in the previous experiment where the
AF/F interface was present during irradiation (cf. Fig.~\ref{Insitu},
circles). Only a slight reduction of the bias field is observed.

This result leads to the conclusion that the F layer has to be
present during irradiation for the enhancement mechanism to work.
The question remains how the F layer affects the modification
process. A pure thermal effect has already been ruled out by the
fact that the enhancement of the shift field remains after an
annealing process in opposite field direction \cite{PRB}. Thus,
only strucural changes remain possible. The F layer can cause
elastic stress/strain that influences the interaction of the atoms
with the ions. Another reason could be that the magnetic
interaction between both layers is required. In order to answer
these questions, a third experiment was carried out.

\begin{figure}
\onefigure{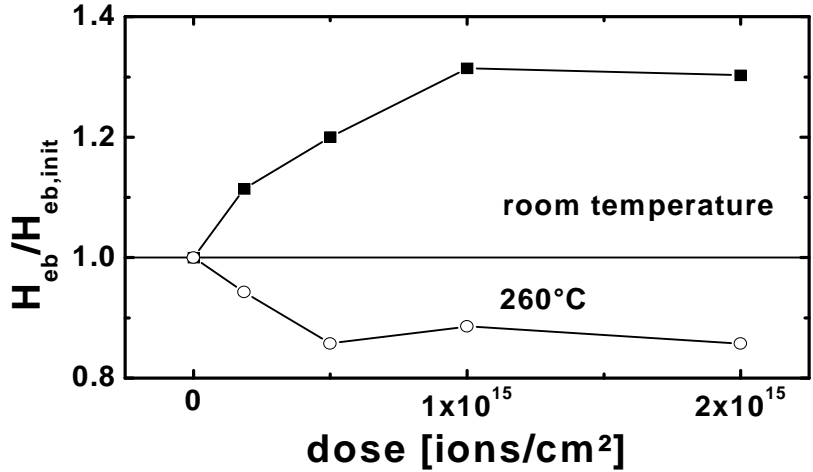}
\caption{Dose dependence of H$_{eb}$ for a sample irradiated
after complete preparation. The data represented by closed
squares has been obtained by irradiating at room temperature, open
circles correspond to irradiation at 260$^{\circ}$C.} \label{Hot}
\end{figure}

A sample with the F layer deposited first was grown completely in one step and
annealed afterwards. First, half of the sample area was irradiated at room
temperature. Second, the other area of the sample was irradiated in an applied
field at an temperature of 260$^\circ$C. This temperature is above the
N{\'e}el temperature of FeMn and thus the coupling between the AF spins is
eliminated, whereas elastic forces, if existent, will persist. Afterwards, the
whole sample was cooled in the same applied field to initialize the exchange
bias effect. The resulting dose dependencies of H$_{eb}$ for both cases are
shown in Fig.~\ref{Hot}. Only the part of the sample which was irradiated with
the antiferromagnetic order present shows an enhancement of the exchange bias
field. The areas that were irradiated at 260$^\circ$C only show a decay.
Therefore elastic strains can be ruled out as the origin of the discussed
enhancement mechanism.

\section{Discussion}
The mechanism for reduction and suppression of the exchange bias
effect by ion irradiation in the NiFe/FeMn system has clearly been
identified to be caused by interactions of the ions with the atoms
in the vicinity of the F/AF interface. This effect can be
attributed to interface mixing. The modifications that cause the
enhancement of the bias field take place in the volume of the AF
layer. These findings are consistent with the models proposed in
Refs.~\cite{PRB,Milt}.

Following the assumptions of Refs.~\cite{PRB,Milt}, one expects a
strong enhancement of the exchange bias field for the case of the
latter experiment. Due to the induced defects, after a
field-cooling procedure the AF layer should develop more domains
compared to the as-prepared case and therefore exhibit stronger
bias, especially as the suppression effect of interface mixing is
avoided in this experiment. The F layer is not involved in this
scenario at all. Yet, this behavior is not observed experimentally
(see Fig.~\ref{Insitu}). Therefore, we conclude that the F layer
has to be present during irradiation. The last experiment shows
unambiguously, that the exchange bias enhancement is not due to
potential elastic effects caused by the presence of the F layer.
In contrast, we conclude that both the magnetic order in the AF
layer and the exchange coupling to the F layer are necessary for
the enhancement mechanism to work.

Our results can be understood in terms of the proposed model if one assumes
that the placement and/or type of defects is not completely random but to some
extent steered by the anisotropic forces in the AF. In the creation process of
a pinning site it arranges in a way, that minimizes the total energy of the
system. Because of the exchange coupling this also includes the energy of the
F layer in a field. Since in all experiments the F layer was saturated either
by the internal bias field or an external field, the arrangement of the
defects will take place in a fashion where the total energy is minimized
resulting in a higher shift field \cite{PRB}. If the spin system of the AF
layer is disordered during irradiation or, equivalently, the F layer is
missing, defect placement occurs randomly. In the first case the disordered AF
layer fails to couple to the saturated F layer providing the preferred
direction. In the second case, there is no preferred direction. In both cases
no enhancement is observed. This data fits to recent simulations where defects
were modeled as local enhancement of the uniaxial anisotropy of the AF layer
\cite{Kim01}. An enhancement of the bias field was only found when the
anisotropy axes are aligned. Defects with random anisotropy axes caused a
reduction of the bias field. The relevance of the magnetic forces is
especially intriguing as the amount of energy placed into the layers by the
ions (i.e. several eV/monolayer according to SRIM simulations\cite{TRIM}) is
large compared to the magnetic energies involved. The exact mechanism of the
interactions between ions and target atoms in these structures needs to be
determined.

In conclusion, we have shown that the observed enhancement of the
exchange bias field in irradiated NiFe/FeMn bilayers is caused by
modifications of the bulk AF layer. It cannot be explained in
terms of a pure statistical domain size argument, and in this
point it differs from the phenomena observed in diluted
antiferromagnets. The influence of the magnetic exchange force
across the F/AF interface has to be taken into account. The
suppression of the exchange bias effect is attributed to changes
at the F/AF interface.

\acknowledgements Fruitful discussions with R.~L.~Stamps and
M.~D.~Stiles are gratefully acknowledged. We thank D.~McGrouther
and S.~Blomeier for valuable help with the TEM studies. The work
is supported in part by the European Communities Human Potential
programme under contract number HPRN-CT-2002-00296 NEXBIAS and by
the Deutsche Forschungsgemeinschaft.


\begin{thebibliography}{0}


\bibitem{Me56}
  \Name{Meiklejohn W. H. \and Bean C. P.}
  \REVIEW{Phys. Rev.}{102}{1956}{1413}.

\bibitem{Mal87}
  \Name{Malozemoff A. P.}
  \REVIEW{Phys. Rev. B}{35}{1987}{3679}.

\bibitem{Koon99}
  \Name{Koon N. C.}
  \REVIEW{Phys. Rev. Lett.}{78}{1997}{4865}.

\bibitem{Sc99}
  \Name{Schulthess T. C. \and Butler W. H.}
  \REVIEW{J. Appl. Phys.}{85}{1999}{5510}.

\bibitem{Ki99}
  \Name{Kiwi M., Meija-Lopez J., Portugal R.D \and Ramirez R.}
  \REVIEW{Appl. Phys. Lett.}{79}{1999}{3995}.

\bibitem{Now02}
  \Name{Nowak U., Usadel K. D., Keller J., Milt\'enyi P., Beschoten B. \and G\"untherodt G.}
  \REVIEW{Phys. Rev. B}{66}{2002}{014430}.

\bibitem{Milt}
  \Name{Milt\'enyi P., Gierlings M., Keller J., Beschoten B., G\"untherodt G., Nowak U. \and Usadel K. D.}
  \REVIEW{Phys. Rev. Lett.}{84}{2000}{4224}.

\bibitem{Shi}
  \Name{Shi H., Lederman D. \and Fullerton E.}
  \REVIEW{J. Appl. Phys.}{91}{2002}{7763}.

\bibitem{PRB}
  \Name{Mougin A., Mewes T., Jung M., Engel D., Ehresmann A., Schmoranzer H., Fassbender J. \and Hilebrands B.}
  \REVIEW{Phys. Rev. B}{63}{2001}{060409R}.

\bibitem{TRIM}
  \Name{Ziegler J. F., Biersack J. P. \and Littmark U.}
  \Book{The Stopping and Range of Ions in Solids}
  \Publ{Pergamon, New York, Oxford}
  \Year{1985}.

\bibitem{Woods}
  \Name{Woods S. I., Ingvarson S., Kirtley J. R., Hamann H. F. \and Koch R. H.}
  \REVIEW{Appl. Phys. Lett.}{81}{2002}{1267}.

\bibitem{Kim01}
  \Name{Kim J. V. \and Stamps R. L.}
  \REVIEW{Appl. Phys. Lett.}{79}{2001}{2785}.




\end{thebibliography}
\end{document}